# Key role of work hardening in superconductivity/superfluidity, heat conductivity and ultimate strain increase, evolution, cancer, aging and other phase transitions

V. P. Kisel[*],

*Institute of Solid State Physics, RAS, 142432 Chernogolovka, Moscow district, Russia*

The shear/laminar flow of liquids/gas/plasma/biological cells (BC), etc. is equivalent to dislocation-like shear of solids. The turbulent flow is the next stage of deformation/ multiplication of dislocation-like defects and their ordering in sub-grains and grain-boundaries, then grains slip-rotation in the direction approximately perpendicular to the shear flow. It is shown that phase transitions are governed by unified deformation hardening/softening under hydrostatic pressure, particle irradiation and impurity (isotope) chemical pressure, hard confining conditions and cooling, etc. thus changing electric, magnetic, ferroelectric, thermal, optical properties[1-2]. Dislocation-like work hardening, DWH, is determined by non-monotonous properties of dislocation double edge-cross-jog slip, and ultrastrong DWH gives the lowest drag for any dislocation-like plasticity at phase transitions. This provides the same micromechanisms of the ultimate stage of conventional deformation (superfluidity) of ordinary liquids, i.e., water, kerosene and glycerin, liquid and solid He, quasi-particle condensates. The key role of DWH is confirmed for superconductivity, integer and fractional quantum Hall effects and the enhancement of ultimate strain and diffusion under deformation down to nanostructures, etc. Phase transformations in biological cells (explosive events of diversity and population of species and diseases – for example, locust and plaque bacteria, evolution, aging and cancer[2], bursts in the development of human intellectual possibilities (languages, culture, arts and sciences, history, etc.) depend on the same deformation effects in biological evolution.

## 1. INTRODUCTION

Up to now a unified model of superconductivity, superfluidity, supersolidity, deformation, diffusion, cancer, etc. remains a key issue. Meanwhile new experiments give more and more confirmations that there is much in common between these effects and ordinary deformation. Since the work[3] the papers[4-8a] gave the comprehensive evidence for the key role of deformation under phase transitions in various materials. Numerous new data confirm this too, for example: the correlation between shear elastic modulus and glass transition temperatures in many metallic glasses[5b] (and see below). The initial stages of deformation of solids, liquids and gases (laminar flow) are structurally identical to the primary straight slips in early deformation stage of solids. The next deformation stage usually meets a higher DWH of flow with decreasing size of cells, grains, and their intensive rotation[9] in the direction perpendicular to the sense of shear flow – the so-called grain-boundary sliding in solids or critical velocity for the beginning of turbulent flow in liquids. The origins of these ones correspond to the appropriate threshold strains-stresses. Literature data offer dislocation-like mechanisms of plastic flow in solids/glasses/ liquids/solid gels/solid polymers/biological tissues, in which the so-called viscous flow equals dislocation shear: in solids the so-called viscosity is equal to the reversed value of DWH of crystals and it is very sensitive to the parameters of deformation: type of crystals, flow rate, temperature, impurity content and state, confining effects, etc.[9,10]. The increase of non-monotonous crystal DWH arrests the primary mobile dislocations, therefore the dislocation sources of primary and then of the other slip systems activate different slip/laminar systems of deformation curve in solids, liquids/plasma /gases up to their turbulent flow, thus increasing the ultimate length of primary stages in solids, liquids, polymers, etc.[10,11a,b].

## 2. THE RISE OF DIFFUSION AND ULTIMATE STRAINS IN DWH STRUCTURES.

A.H. Cottrell found the similarity between the laminar/turbulent flow in liquids and plastic flow in solids[12] which was confirmed later at high stress rates of deformation[10]. The work[13] was the first to point out that the voltage-current dependences in superconducting state are suggestive of the standard stress-strain deformation curves in solids. There is a lot of evidence for the deformation origin of current flow and all phase transitions – in magnetic (including flux-line lattices and phase transitions under magnetic field **B**), ferroelectric, electric (including corrosion and phase transitions under current), chemical (catalysis, dissolving, deposition, chemical reactions, mechanochemistry, etc.), thermal (melting, crystallization, sublimation, condensation, thermal conductivity[14]), mechanical (sample irradiation with particles, diffusion, mechanochemical reactions, etc.) and biological (activation from DNA up to cell differentiation, proliferation and growth, diseases, cancer and aging) transitions[1-10,15-17]: electroplastic effect[18-19], diffusion/particle-irradiation and ion electromigration or electrodeposition[20], torsional oscillators in the medium change the internal stresses and noticeably deform the samples[21-22], etc. Therefore the same low-temperature athermal behavior of diffusion[23] and crystallization (especially in the medium with ultralow density)[22,24] or in chemical reactions[25], thermal conductivity[26] and plastic flow under superconducting transition[27], etc. was due to classical (non-quantum) deformation mechanisms[14,17].

It is worth noting that the scaling of stresses for plastic flow from the deformation $\varepsilon = 10^{-8}$ (internal friction) up to $\varepsilon = 0.05$[4-10,17,28] and more strictly confirms the same micro-mechanisms of flow in different materials along the stress-strain curves up to nanostructures ($\varepsilon \sim 10^3$) and fracture[5,29,39]. The same effects of jump-wise flow of deformation in nanostructures[30], flux line lattices[58] and magnetic resistance[36b] in superconductors, in current flow in solid He[31] and velocity of micro-fluidic flow[32], etc. corroborate this (see below).

This is in line with an abrupt enhancement of primary stage slip length[11] (Figs 12a,33), ultimate strain of nanostructures[30], drop the effective viscosity for motion of particles/ dislocations in ordinary crystals[10,33], hyperdiffusion under lower temperature[34] which depend on crystal type and orientation, strain rate and the frequency of oscillating loadings, temperature, type, state, size and concentration of impurities[2,11,17], heat prehistory, cooling/heating hysteresis, hydrostatic pressure, irradiation, content of isotopes, lattice symmetry, magnetic and electric fields, "size-effects", orientation angle anisotropy and the structural properties of substrates and films, confining and frequency effects, etc. These data and scaling of flow stresses in various materials undisputably point to the governing role of double cross slip of dislocation-like defects in deformation localization of DWH of materials[3-10,17,28].

## 3. SUPERCONDUCTIVITY

The moving wires/particles in solids (mechanisms of deformation of matter under atomic-scaled particles, torsional oscillators or macroscopic rods are the same along the deformation curve[4-10,17,28, 29,39]) macroscopically deform the matrix[1,21,22], and their motion through the crystal lattice can be well described by dislocation motion, multiplication, work hardening and softening[1,4-11]. The dislocation lines and bands become thinner and shorter (localization of deformation) with temperature reduction and increasing DWH due to dislocation-like double edge-jog cross slip (DDEJCS)[4-8a,b,17,35], localize in the primary-slip-stage directions at the beginning of DWH[17] and in superconductors/insulators they aggregate the charges[36a,b] to the ordered superstructures (stripes, charge-density wave, CDW)[37] (Fig.1b) due to their arrest at lattice defects[9,39]. The effective width of these stripes decreases monotonically under increasing DWH of current[40] like the slip bands under DWH of solids[4,17,]. The structure of Zn single crystals in normal and superconducting states (Fig. 11 in [40]) vindicates that single electrons (normal state) and Coopers' pairs (superconducting state) noticeably change the DWH and the heights of DDEJCS. This means that charges are closely interrelated to localized DDEJCS around the charges (compare with the so-called phonons in BCS theory[14,17])[1,8,10] and jog motion with entrapped charges determines electrical conductivity $\sigma$ and dislocation mobility -crystal viscosity[10,39]. This is in line with the set of works which discovered the same influence of various factors (stage of deformation, pre-deformation and strain, impurity and superconducting phase concentrations, state, size and type of impurity clusters and precipitates, stress rate, heat treatment, temperature, etc.) on crystal softening under superconducting transition and crystal softening under ordinary deformation [27,39,41].

Note that low-amplitude ($10^{-2}$-$10^{-1}$MPa) drops of the flow stress in Pb-In alloys at ~3.5K

appeared in a restricted concentration range of In (~20 to 55 at.%) like it is in the ordinary alloys, but in the superconducting state only[42-43]. DDEJCS at the impurity clusters depends on their size[27,42-43] and determines the DWH/DWS(dislocation-like work softening)[3-10,17,27] here in cooperation with DDEJCS at single and superconducting electrons (Coopers' pairs), i.e. the standard jogs coincide with the superconducting jogs, so the combined mobile jogs can drag the mobile charged pairs along the dislocations up to the obstacles (see part 4a) thus forming discontinuous ordered structures of charged aggregates.

These charged supermodulations are mainly along the **a**, **b** and **c** axes, and what's more the periodicity along the **a** axis is less than along the **b** axis in $Bi_2Sr_2CaCu_2O_{8+x}$ at 300K[44]. In terms of deformation bands this means the higher internal stresses, thinner slip bands and higher conductivity along the **a** axis. In single crystals CuO (the basic compound for high-$T_c$ cuprates) charge-ordered domains and normal-lattice domains exist alternatively and the resistances $\rho_{\|a} \leq \rho_{\|b}$ (T=100-300K), but $\rho_{\|c}$ is more than 1 order of magnitude lower than $\rho_{\|a}$ and $\rho_{\|b}$ [37]. The anisotropic charge transport is consistent with the real-space image where the charge stripes run along the hard **c**-direction[37]. This current anisotropy is supported by the same anisotropy for dislocation mobility and multiplication in the so-called hard or soft secondary planes for dislocation double cross slip: if cross slip plane is hard (**c**-direction), the heights of jogs is lower, dislocation bands are thinner and dislocation mobility[45] and conductivity in primary plane (**a-b**) is high[37] and *vice versa*. The double jogs and kinks originated from dislocation double cross slip near the charges/defects then slip freely along the dislocations up to various stoppers[39], dragging them along [39], decreasing the $\rho_{\|c}$ and forming the charge stripes run[37,46] (see parts 1-4). The so-called quantum critical point is simply the threshold strain (or applied field $E_{th}$) to trigger them[47]. Dynamics of CDW in $NbSe_3$ at low temperatures have the same features as the mobility of single dislocations and their slip bands: the existence of threshold electric field $E_{th}$ to start CDW moving which increases with decreasing temperature, a current density **j** is proportional to the *nonlinear* CDW sliding velocity, the impurities cause the CDW to oscillating (jump-wise) motion in high-quality crystals, the frequency of oscillations rises sharply with applied field E growth, $\mathbf{j}(E>E_{th})$ is activated in temperature and increases exponentially with E to more nearly temperature-independent value and often by abrupt, hysteretic "switch"[34,39] which depends on concentration of pinning defects, temperature, etc.[39], the **j** of CDW is orders of magnitude smaller than **j** for single particles, but in the low-velocity branch the activation energies for single-particle conductivity and the CDW conductivity are both comparable to the CDW energy gap (compare with the scaling of mobility of individual dislocations and slip bands in[17]), the $E_{th} \sim 1/d$ for crystal thickness $d < \sim 20\mu m$ (deformation size effect)[38], the dependence on the current-density-sweep rate $d\mathbf{j}/dt$ (compare with the data of [4,35,48]) and grain size, etc. The properties of quasiperiodic ionic-current bursts during steady-state current in strained crystals of solid $^4He$ [31] corroborate the general deformation origin of step-wise flow of charged particles. Therefore it is worth noting that the comparison of works[36a,46] obviously shows that: the first charge aggregation in YBCO takes place at temperatures $T^* \sim 200K > T_c$ (still in the normal state), whereas Coopers' pairs formation starts only at $T \leq T_c$ (the border of superconducting state)[36a]. Much smaller magnitude of the fluctuations in concentration of the itinerant (not bounded) holes in the **a**-direction as compared to the **b**-direction[36a] means the thinner slip lines with charges, the lower heights of edge jogs and kinks, higher internal stresses and local deformations in the first case, the lower resistance in this direction. The appearance of higher internal friction hysteresis (compare with the data of [34]) at $T^*$, $T_c$ [46] means the movement of rare and jogs/kinks/charges with higher heights in the first case and the faster and lower jogs-kinks of higher density due to higher DWH[17,19] in the second one (this conclusion is confirmed by the data of [49]). Then mobile jogs-kinks with charges gradually come to arrest and then to their motion-multiplication[9], thus forming discontinuous charge aggregates (at $T^* > T_c$). At lower $T < T_c$ the jogs-kinks-charges with the gradually lowering sizes (down to the elemental size) and progressively growing density [35,39] and conservative slip velocity decrease the hysteresis, but increase crystal conductivity in this direction. If at $T_c < T \leq T^*$ (normal state) the so-called anisotropic pseudo-gap originates in the excitation spectrum due to charge aggregates at dislocation rare high jogs-kinks in discrete slip bands, at $T \leq T_c$ the standard energy gap forms with the highest density of jogs-kinks-Coopers' pairs trains along dislocations within dense slip lines (the BCS coherent state of Cooper pairs condensate). It is worth noting that only dislocation double cross slip localization at low temperatures[4,8,10,17,35] induces the aggregation of charged particles in Coopers' pairs in complete agreement with Pauli principle. Again everybody can see that superconductivity and superfluidity of particles flow are determined by the same dislocation mechanisms- free movement of conservative jogs - one-side kinks mainly along the straightened screw dislocations in work-hardened crystals[17,39], and this is confirmed by the scaling of the stripes/conductivity energy gaps and activation energy for dislocation-like motion and diffusivity[38,50a,b,51a], observation of super long–range correlations up to the micrometer scale in sliding CDW[51b,9]. The conducting two-dimensional structures (electron layers, Wigner crystal, graphene) with the ordered[52] subsurface layer of substrate may depend on the restrictive geometry (the proximity effect[8a,b]) and conducting properties of thin 3-dimensional layer with standard deformation properties under current flow[53].

It is well known that any type of phase transitions in all classes of materials is usually dependent on and is accompanied by closely related physical (plastic deformation, heat-generating/absorbing effects, origin of various defects, nucleus and clusters of new structures, acoustic and electromagnetic radiation over extra-wide ranges of frequencies, charged and neutral particles emission in nuclear decay and fission, structural transformations, etc.), chemical (new atomic-molecular and electron configurations, valence states, spin orientations, changes of bond lengths and angles, etc.) and biological (broken atomic and molecular bonds, specific defects of metabolism - reactive oxygen species (ROS), conformation or structural changes of biological forms, DNA activation, cell proliferation and growth, etc.) transformations in parallel. So, it is these features that may be the distinguishing signs of the key role of plastic deformation at any phase transitions in solids, liquids, plasma, gases and BC. For example: generating/absorbing effects of electromagnetic and acoustical radiations at chemical reactions, expanding ultracold plasma, normal-to-superconducting transitions, cell mitosis and during intensive relative motion of ordinary liquids, normal component with respect to the superfluid one in the liquid $^4$He-II, spontaneous heat-generating/absorbing and electromagnetic field effects at melting, crystallization, etc., the appearance of superfluid vortices in $^4$He following rapid quenching to the superfluid state, vortex formation by merging of initially indeterminate relative phases of Bose-Einstein condensate or during its condensation, deformation effects at phase separation in Fermi gas; rapid isotropic to nematic transitions produce topological defects in nematic liquid crystal.

Alternating loading of crystals leads to their softening[39] or hardening[29] depending on the parameters of samples (grain sizes, type, size and concentration of impurity phase, thermal predeformation and concentration of lattice defects, etc.), frequency, amplitude, temperature and strain rate of deformation tests. The same sample hardening/softening exists at their simultaneous loadings with flow stress $\tau$ slightly below or above the yield stress $\tau_y$ and the ultrasonic stresses with much lower $\tau$ (F. Blaha - B. Langenecker effect)[39,54]. The work[55] demonstrates the obvious hardening/softening effect (the Bauschinger's effect is determined by the same properties as the DDEJCS in narrow ranges of temperatures, impurity sizes, types and concentrations, sample DWH/DWS, strain rates, etc.[39] – compare with parts 4a,b of low-temperature loading of 2-dimensional conductor with electric and magnetic fields.. These facts are in line with the temperature of superconducting transition $T_c$ shift under ordinary sample work hardening or softening due to simultaneous effects of constant and high-frequency lower currents[56] or under constant and weak alternating magnetic field $B^{57}$ in comparison with the data of only one loading factor, step-wise deformation of flux line lattices[58] or magnetoresistance[36b] in superconductors and nearby the melting point[59], etc.

So the current-voltage plots are analogous to the stress-strain curves where the elastic Hook's law is identical to the Ohm's law for electric current, the influence of magnetic field on the plastic properties of solids, liquids, etc. (DWH/DWS of the so-called magnetoplastic effect, MPE[8a,60a,b]) can suppress[42] or promote[61] the unstable deformation of sample like the super-conducting state[42,43]. The MPE explains the effects of orientation of electromagnetic field in comparison with the direction of current flow, its amplitude and frequency influence on normal conductivity or superconducting (superfluid) states, phase transitions, and so on.

The deformation approach to phase transitions explains the mysterious vanishing of isotope shift in the temperature $T_c$ for optimal doping in cuprates, and much bigger shift at low doping levels. This effect can be understood in terms of DWH/DWS of internal stresses $\sigma_i$ in hard materials: at optimal doping the $\sigma_i$ is larger than its change $\Delta\sigma_i$ at replacing the heavier ions with the lighter ones due to the lower radius of heavier ions and even the lattice parameter is often less than the same of lighter ones, but in the case of low doping levels the so-called chemical pressure $\sigma_i$ is comparable with $\Delta\sigma_i$ at the same replacing and this variation is more observable at a change of $T_c$. So it is quite natural that the switching the iron isotopes in non-superconducting samarium and barium compounds also increases the temperature of magnetic transition (like the ferromagnetic and other phase transitions) as in the new superconducting iron-and-arsenic compounds at the same changes of DWH deformation conditions. This explains all previous effects at phase transitions and confirms that the superconductivity is only an ultimate stage of ordinary deformation which governs all phase transitions.

## 4a. INTEGER QUANTUM HALL EFFECT

In much work-hardened two-dimensional electron gas confined between two hard semiconductors at high magnetic field B (>1T) there is the *integer* quantum Hall effect, zeros in the longitudinal (*dissipative*) resistance $R_{xx}(B,T)$ in between the increasing/decreasing jump-wise bursts while the *transverse* (Hall) electrical resistance $R_{xy}(B,T)$ is accompanied by intermittent step-wise gradual raises and plateaus $R_{xy}(B)$=const with increasing their lengths and decreasing frequency under B rise. In the case of $R_{xx}(B,T)$ its amplitude of jumps is increased or decreased as B is raised, but it decreases with temperature drop and its frequency always reduces with B growth like for $R_{xy}(B)$. The threshold $B_{th}$ for the beginning of jump-like resistance in quantum Hall effect increases with temperature reduction[62]. The same results are well known for ordinary step-wise dislocation-like plasticity of solids, water, etc.[15,42,43,59-61,63], deformation nanostructures[29a,b,c,30], solid He[31], crystal growth, magnetic flux line lattices[58] and magnetoresistance[36b], where B plays the role of softening/hardening stress in the so-called non-monotonous MPE [60a,61,63a,b], for periodical explosive bursts of life diversity and population during biological evolution, etc. Amplitudes of standard step-wise plastic flow and their frequencies depend on the same parameters and show the same non-monotonous dome-shaped dependences *vs* DWH like the amplitudes of current bursts under quantum Hall effect and temperature of superconducting transition $T_c(P, c, \sigma, \varepsilon, Ø,$ etc.), where P is hydrostatic pressure, c is concentration of various type and state of impurities, $\sigma$ is deformation stress and $\varepsilon$ strain, $Ø$ is grain size. The fine changes in structure of individual dislocations determine micro- and macroscopic DWH and DWS effects[3-8a,b,,17,23,39,63a,b] in samples due to:

i) dislocation athermal bowing between the obstacles (DWS) and ii) dislocation double cross-edge-jog slip (DDCEJS) or double-edge-kink slip (DEKS) at the lattice defects under the dislocation line tension (DWH-DWS), iii) the *dissipative* non-conservative jogs climb in the direction of applied force (DWH) and iiii) the conservative slip of edge jogs and kinks along the screw dislocations in transverse direction to the applied force-current (DWS) [4,7,9,15,17,39,63]. The electron interaction with DDCEJS –DEKS[8a,39] is confirmed by the registration of electric bursts in *longitudinal* and with much less amplitudes in *transverse* directions to the mobile dislocation twinning under *active macroscopic deformation*[64a,b] and the reversible effect: high electric current sets in motion *edge* dislocations only[19] (compare with work[18]) because screw dislocations in crystals with DWH are strongly pinned by jogs[39]. Note that the deformation of Nb single crystals in normal and superconducting states gives approximately the same rise of amplitudes of electrical bursts under dislocation twinning in *longitudinal* direction with the growth of total deformation $\varepsilon(\tau)$, while the amplitudes of *rarer bursts* in *transverse* direction are *with much less*

*amplitudes* of electrical currents[64a] (Fig.3) and they do not depend on ε (Fig. 4). These differences are in line with the scaling of mean pathlengths of mobile dislocations and their mean number[9] and it may be concerned with the fact that dislocation twinning at DWH is determined by the strong pinning of trailing screw partial dislocations by numerous climbing jogs of noticeable heights due to intensive double cross slip of long dislocation loops at numerous defects under macroscopic deformation[39]. This prevents the effective jogs-kinks conservative slip along them and increases the forced entrainment of charges in *longitudinal dissipative* direction[64b]. The reality of this mechanism is confirmed by the emergence of intrinsic impurity and point defects on sample surface or defect accumulation after dislocations in the bulk due to conservative and non-conservative glide of edge jogs and kinks with impurity entrainment[39]. But this pinning process is much weakened under lower strains-stresses in front of short mobile charges, and this is in line with the fact that the $T_c$ is usually higher under moderate deformation.

So we consider that the increase of *dissipative* $R_{xx}$ bursts (it means the DWS of crystal in this direction) is concerned with the decrease of ii) DDCEJS/DEKS and iii) slow dislocation motion by climb between obstacles (moderate DWH), while the numerous spontaneous jog-slip jumps provide the *transverse* step-wise changes of $R_{xy}$ (DWS) determined by the mechanisms ii), i) and iiii). They govern the plateaus lengths which are increased with **B** rise due to gradual crystal softening and depend on **B** orientation. The fact that the minima of $R_{xx}(\mathbf{B})$ coexist with plateaus in increasing values of $R_{xy}(\mathbf{B})$ in terms of the above dislocation mechanisms of conductivity means that in these cases the crystal *longitudinal* direction is in moderate DWH state and dislocations do not noticeably bow out between the obstacles, *dissipative* climb of non-conservative jogs and slip of kinks provide slow deformation processes only (the quasi-arrested state for the *longitudinal* direction). This means that dislocations are straightened mainly along directions of the slip Burgers vectors **b**, and the most of screw dislocation conservative jogs/kink-charges take "a train shot" (fast deformation process) along them in *transverse* direction to the *longitudinal* one[17,19,39] up to the long-term pinning obstacles (the rise of $R_{xx}(\mathbf{B})$ is replaced by the decrease of $R_{xx}(\mathbf{B})$ due to jog-kink pinning and the amplitude of these cycles is increased with facilitation of jog-kink unpinning in the direction of gradual softening). The increasing amplitudes of $R_{xy}(\mathbf{B})$ steps at $R_{xx}(\mathbf{B})$ bursts is concerned with somewhat growth of *dissipative* motion of jogs and slow deformation with kinks, the increasing lengths of plateaus with keeping constant $R_{xy}(\mathbf{B})$ is due to gradual increasing time of intermittent retardations of dislocations with climbing jogs (moderate DWH). The lattice defects gradually retard the jogs-kinks motion, but the rise of **B** makes the crystal softer and increases the amplitudes of the $R_{xx}(\mathbf{B})$ bursts and $R_{xy}(\mathbf{B})$ steps.

If the retardation of jog-kink slip is larger at low softening effect of **B**, and it increases faster in comparison with crystal softening at gradual low **B** rise, the amplitudes of $R_{xx}(\mathbf{B})$ bursts progressively decrease due to the origin of denser and lower-heights jogs-kinks with their increased *transverse* runs-currents. And this time-lengthening DWH promotes the same lengthening plateaus of minimal $R_{xx}(\mathbf{B})$ and the rise of $R_{xy}(\mathbf{B})$ with **B** in larger steps and lengths of plateaus in consecutive order. It is worth stressing that the decrease in $R_{xx}(\mathbf{B})$ bursts is typical for *low* **B** softening ranges.

It was shown earlier that the increasing jumps in crystal softening at MPE are accompanied with decreasing whole numbers of electron **g**-factors at electron paramagnetic resonance[60a]. These whole numbers of **g** are supposed to be determined by the moderate DWH at ultimate dislocation pathlengths and appropriate decreasing sizes of climbing jogs which are multiple to lattice parameter and thus decreasing the density of charged vacancies per jog-kink after them[15,63a,b]. This tendency is in line with the data on quantum Hall effect at **B** increase, where

$$R_{xy} = h/i\mathbf{e}^2, \quad (1)$$

so the decreasing integer **i** plays the role of *effective charged-particle density per jog-kink* (the whole number of filling factor **i**), h is Planck's constant and **e** the electron charge.

**4b. FRACTIONAL QUANTUM HALL EFFECT.**

Like an *integer* quantum Hall effect the *fractional* quantum Hall effect shows the plateaus in the

*transverse* Hall resistivity $R_{xy}(\mathbf{B})$ which coexist with deep values of *longitudinal* $R_{xx}(\mathbf{B})$ down to zeros, the values of plateaus-steps in $R_{xy}(\mathbf{B})$ and the jump-like bursts of $R_{xx}(\mathbf{B})$ increase with $\mathbf{B}$ growth, but the magnitudes of fractional filling factor $\upsilon$ are much lower than **i** ($\upsilon = \mathbf{r}/\mathbf{p}$, $\mathbf{p} = 1, 3, 5,…$- the odd whole numbers only, while the **r** is any whole number). The fact that the minima of $R_{xx}(\mathbf{B})$ coexist with plateaus of constant $R_{xy}(\mathbf{B})$ with increasing values of $R_{xy}(\mathbf{B})$ in the frames of dislocation mechanisms of conductivity mean that the minimal stages of *longitudinal* $R_{xx}$ are in accord with moderate DWH of crystal state where dislocations do not markedly bow out between the obstacles, the *dissipative* process of jogs climb and slow slip of kinks provide slow plastic flow. This means that dislocations are straightened mainly along the directions of the slip Burgers vectors **b** thus making easier dislocation jogs/kink-charges "train shot" along them (fast deformation process) in *transverse* direction to the longitudinal one[17,19,39] and reducing the $R_{xx}(\mathbf{B})$, keeping $R_{xy}(\mathbf{B})$ constant at the plateaus and gradual increasing $R_{xy}(\mathbf{B})$. When the lattice defects noticeably retard the jogs-kinks motion, the rise of **B** makes crystal softer with gradual increase in resistances: the larger amplitudes of $R_{xx}(\mathbf{B})$ bursts and $R_{xy}(\mathbf{B})$ steps with expanding parts of the $R_{xy}(\mathbf{B})$ minimal values and the plateaus of constant $R_{xy}(\mathbf{B})$ in between.

These hardening and softening mechanisms are closely interdependent and interchangeable, and this is confirmed by the data of work[65] which demonstrate that the changes in carrier mobility transfer integer quantum Hall effect into the fractional quantum Hall effect. This is absolutely natural in the frames of dislocation model with jogs and kinks for current flow.

It is well known that usually the increase of resistance is related to the sample DWS[60a]. This softening effect is increased with **B** rising at MPE[60a,b,61] and corresponds to a decrease of the whole numbers of entrained charge-particles per mobile jogs/kinks (any whole numbers of numerator **r**) and increase the odd whole numbers **p** of elemental conservative jogs-kinks per jogs-kinks unit length [4,9,10,15,17,39] in the denominator of the fractional number of filling factor $\upsilon$ for current bursts[62]. It is worth stressing that a vacancy-hole is equal to double elementary jogs (conservative and non-conservative ones) therefore any number of Coopers' pairs ("train shot") always gives even a number of conservative jogs - one-side-motion kinks plus one conservative carrier jog-kink. So the density **p** of conservative jogs-kinks is always determined by the odd whole number, and the $\upsilon$ has the same meaning as **i** : the *effective charged-particle density per jog-kink* (the fractional number of filling factor $\upsilon$). So the fractional quantum Hall effect has the same mechanism as the integer quantum Hall effect and they are determined by the step-wise decreasing effective charge particle density per jog-kink while it is usually quasi-continuous in the superconducting one with **B** rise.

The features of charge ions movement in hexagonal phase of solid He were explained by the motion of dislocation jogs in cross slip planes and kinks in slip ones too[7].

The authors of the work[62] have found a linear correlation between the value of $V_{cell} \cdot \mathbf{B}/eR_{xy}$ (where $V_{cell}$ is unit cell volume) of ordinary Hall effect and superconducting transition temperature $T_c$ in $Bi_2Sr_{2-x}La_xCuO_{6+\delta}$ under different hole doping (Fig. 3):

$$T_c \sim V_{cell} \cdot \mathbf{B}/eR_{xy} \qquad (2)$$

This scaling gives the correlation $T_c \sim \sigma_{xy}=1/R_{xy}$ at constant **B** which indisputely means that the superconductivity is determined by the charge (Hall) conductivity in transverse direction to the basic current flow thus corroborating the side conservative motion of jogs/kinks along the dislocations in front of moving charges whereas the non-conservative jog climb and kink slip provides dissipative $R_{xx}$. These conclusions are in line with the fact that increasing mobility of charges suppresses the integer quantum Hall effect and promotes the fractional quantum Hall effect[65]. And what is more interesting: if the total energy E of charged particle is E = h**v**, where **v** is the frequency of particle oscillation then the particle velocity **v** approximately equals

$$\mathbf{v}(T, P, c, \sigma, d\sigma/dt, etc.) = c\mathbf{v} , \qquad (3)$$

where **a, b, c**(T, P, c, $\sigma$, $\varepsilon$, d$\sigma$/dt, etc.) are lattice parameters, $V_{cell} = \mathbf{c} \times \mathbf{a} \times \mathbf{b} = \mathbf{c} \times \mathbf{S}$, where **S** is usual cross section of unit cell volume of charged vacancy-hole, T is temperature, P is hydrostatic pressure, c is concentration of various type and state of impurities, $\sigma$ is deformation stress and $\varepsilon$ strain, d$\sigma$/dt stress rate, so the equations (1-3) give the next expression

$$T_c \sim \mathbf{i}(\text{or } \upsilon)\mathbf{evBS}/E \sim F_L/E, \qquad (4)$$

where $F_L(T, P, c, \sigma, \varepsilon, d\sigma/dt, \text{etc.}) = \mathbf{q}(T, P, c, \sigma, \varepsilon, d\sigma/dt, \text{etc.})\cdot[\mathbf{v}(T, P, c, \sigma, \varepsilon, d\sigma/dt, \text{etc.})\mathbf{xB}]$ is the Lorentz force which helps to move the effective charge density $\mathbf{q} = \mathbf{ie}$ in *transverse* direction, $E(T, P, c, \sigma, \varepsilon, d\sigma/dt, \text{etc.})$ is total particle energy which includes the potential energy of charge bonding with lattice (lattice friction or drag force for the charge motion). The expression (4) is absolutely natural for the mobile charge deformation of lattice: the Lorentz force helps to increase charge mobility in this case like in the case of MPE in water[15,63b] or solids[63a] and increases/decreases (magnetoresistive effect) the temperature of superconducting transition, whereas the growth of potential part of charge energy regulates lattice DWS/DWH for charge motion: at $\mathbf{B} = 0$ the $T_c$ is determined by the deformation properties of DDEJCS and DEKS.

The measurements of ultrafast transport electrons at T = 4K in GaAs confirm the high electron velocities up to ballistic motion and *side valley transfer*[49]. The same slow deformation is valid for two-dimensional structures (electron layers, graphene) with the highest electron mobility in strongly hardening structures due to size and confining effects.

Recent experiments on two-dimensional structures which were conducted *at much lower* values of **B** did not show any noticeable jump-wise behavior of $R_{xx}$ like in the case of quantum Hall effect[62], but if the systems had been subjected *additionally* to high enough microwave radiation intensity of the *right* frequencies, the weak jump-like *longitudinal* resistance $R_{xx}$ appears[66] like the typical jump-wise deformation under superposition of active (in basal plane) and ultrasonic loading (in cross-slip plane)[54].

BCS theory of superconductivity is based on charges paring due to their interaction with lattice oscillations-phonons (oscillating particles in the dislocation approach) and their condensation in coherent state, but our deformation-hardening extended approach unites these and current *dynamical* properties depending on all numerous deformation-dependent effects.

In the parts below we'll discuss the additional proofs and applications of the governing role of deformation in other phase transitions.

## 5. SUPERFLUIDITY OF ORDINARY LIQUIDS, SOLID AND LIQUID HELIUM

The initial stages of plastic flow in solids, liquids and gases (laminar flow) are structurally identical to the primary straight slips in early stages of DWH plastic flow at stresses $\tau \geq \tau_y$ (the yield stress), and each work-hardening stage has the inherent threshold flow stress. The structure of new mode of deformation is characterized with decreasing size of sub-grains and cells and their intensive rotation – the so-called grain-boundary sliding[9] which conforms to the critical velocity for the origin of turbulent flow in liquids, at high $\tau$ in solids, etc. Moreover, the correlation was found between the coverage of adsorbed organic layer (rigid covers have the effect of DWH on plastic flow) and the critical current in thin superconducting Nb films[67], hard environment or confining surfaces on the carrier dynamics in carbon nanotubes[68], etc. (the so-called proximity effect[8]). DWH orders the structure of liquids[52], and hard covers of solids stimulate their DWH. Independent detailed analysis of the data on diffusion and waves of crystallization in solid $^4He$[24] has shown that they are governed by the universal deformation mechanisms of deformation localization or suprfluidity in very low dense and friction matter (ordinary gases, dust plasma[69], liquid[70] and solid $^4He$ and $^3He$, water, kerosene, glycerin[71], Bose-Einstein condensation, ultracold Fermi gases, fermionic atom pairs, etc.)[8b] with DWH at room or low temperatures, in thin confining slots, capillaries, while in conventional materials similar effects are much smaller and appear at ultrahigh DWH and strain rates, molecular confinement and lower temperatures only (these conditions corroborate the conclusions of parts 2,3,4a,b). The drop in the so-called viscosity of NaCl single crystals for high-velocity dislocations[10] and the extra deep penetration of high-velocity iron microparticles into steel target at 77K[33], hyperdiffusion under lower temperature[34] *identified with heterogeneous strain in polymer melts*, accelerated deformation of entangled polymers during squeeze flow in narrow gaps[11b] and the step-wise change of ultimate strain in deformed nanostructures[29,30] (compare with parts 3,4a,b). It is the sharp decrease in edge-jog heights under DWH that promotes the step-wise changes of dislocation structures in hardened crystals and nanostructures: the retarded grain boundaries[9] and dissociated dislocations stacking faults[39] are replaced with mobile sub-grain and grain boundaries[9] thus inducing their coagulation, growth and jump-wise flow under higher stresses[29,30a,b,c].

Note that when the short jet of small (3.4 μm) particles is injected into a plasma cloud of big (9.2 μm) background particles there is the lane formation in complex plasma[68] in the same way like the formation of nanofiber-like impurity precipitates along the laminar flow lines in denser superfluid $^4$He-II. in comparison with spherical microparticles fall in turbulent flow of normal liquid $^4$He-I[69]. These examples are typical for phase transitions: new phase falls out along the deformation slip bands in ordinary solids, liquids, gels, polymers with impurity phase, crystallization phase in amorphous materials, etc.

The universality of deformation approach for phase transitions is confirmed by the same thermomechanical effects in ordinary ice growth on the cooled fine porous ceramic in water (with a high rate of several millimeters per day) due to strong confining ordering and DWH of freezing water in the same way as in superfluid $^4$He-II [72]. The well-known non-linear energy spectrum for thermal excitations in liquid $^4$He-II, where the energy is plotted as a function of momentum (1.1K< T< 2.19K)[73] designates the atomic-scaled drag of He atoms to move in the superfluid $^4$He-II as a function of their displacements under impact of slow neutron particles. In other words this is the ordinal lower-temperature stress-strain curve with the typical upper yield point for the deformation of dense and confined $^4$He-II liquid (T< 2.19K) under the movement of He-atoms in comparison with the smooth and lower stress-strain curve at higher temperature for lower dense and not confined liquid $^4$He-I (4.2K). Again this confirms the above numerous arguments in favor of common deformation mechanisms in solids, liquids, plasma and gases and means that the quasi-linear part of these atomic-scaled deformation curves corresponds to the slight reversible motion of individual dislocation-like defects - the single carriers of plastic slip in liquids (laminar flow)[63b], while the first stage of plastic flow means partially irreversible multiplication of dislocation-like defects (the beginning of turbulent flow in confined liquid HeII and free HeI). This is confirmed by the data on the close interconnection between dislocations and thermal flow in solid helium crystals[8b,14,17], the exhausting explanation of various properties of moving ions, charged particles and electrons in solidified gases in terms of local plastic deformation of matrix in front of them[6,7], the key role of deformation localization in the pairing of charges at low temperatures[8a,b] and by the scaling of the flow stresses up to fracture stresses in crystals from atomic to macroscopic scale lengths from solid helium up to diamond and hard ceramics, metallic glasses, etc. [5,17,28,29,63b]. This agrees with numerous experiments which cannot be accounted for by the Bardeen-Cooper-Schrieffer (BCS) theory[74a,b], etc.

## 6. CANCER, DISEASES, AGING AND EVOLUTION

*May be, the cause [of cancer] is the only one and simple, and we think too complicated, that's why we could not answer this question.* Loren Schwarz (Paris)

Since the mechanisms of deformation in solids, liquids, biological cells (BC), gases, dust plasma are the same[2,15], parts 2-5, we have a well-founded expectation to see the same superfluid-hardening phase transition in cancerous cells. The abrupt hardening and high rate of proliferation in tumor cells[2] are well-known, which makes them immortal like the long-living currents in superconducting state or continuous slipping of superfluid He film on the hard walls. If the abrupt DWH of matrix is sufficiently high for a long time, therefore the crystal-liquid-BC plastic stress relaxation has to be very small.

The deformation processes determine phase transitions and DWH/DWS of fatigue in BC (aging) under body metabolism and physiological stress deformation, confining effects include the biochemical chain of genome and protein dynamics/regulation from DNA up to the cell differentiation, proliferation, growth and signaling[17]. The work[75] shows directly *in vitro* that the old (according to our model – the ones with DWH) human fibroblasts harden the pre-cancerous cell subpopulation thus stimulating its proliferation and then epithelial cells transformation into the malignant ones which promotes the growth of entwined tumors in bare mice. In younger

(softer) fibroblasts the DWH and the appropriate cancerous manifestations are weak. The usual long-term time of malignant tumors growth in very old patients with DWH tissues is due to low mismatch stress $\Delta\tau$ between the robust and malignant BC thus confirming that the origin of cancer transformation is concerned with DWH.

Malignant development is heterogeneous process starting from the random local hardening of BC due to various reasons in the same way as DWH of crystals at plastic deformation. Experiments have shown that the exponential growth of relative size of eight types of entwined malignant tumors as a function of the doubling number **n** of transformed cells is accompanied by the unified bell-shaped concentration dependence of active ROS in cancerous cells in parallel with their maximum growth rate and ROS maximum at $n_o$ = 5.6+/- 0.29 of the doubling cell number **n**, then it drops abruptly to the concentration values of *immature and embryonic cells*[75]. In terms of deformation approach to phase transitions this non-monotonous change in concentration of broken bonds-point defects is the independent and undisputable support of the rise of DWH under tumor growth up to moderate values of work hardening at $n_o$ (the safe state for life) and then gradual growth of DWH of cancerous BC up to the dragless cell proliferation and growth at ultrahigh hardening - "superfluid solid" (death state for BC).

Very fast and strong DWH of cells prevents them from aging processes (like electrons from arrest at superconducting transition) and stimulates dislocation-like defects multiplication[9] thus promoting the long-term cell proliferation up to the final mode of their plastic flow - apoptosis of cells (or crystal fracture); this tumor disintegration is the cause of metastasizing[2,16]. The deformation nature of tumor growth is confirmed by the process of its local temperature increase and the changes of impurity content in tumor serum (like the same changes during the crystal growth), etc.

Malignant tumors have the individual irregular structures which are morphologically different, and, also, are changed according to the independent cells differentiation and growth character *with the features of immature and often embryonic cells*. This correlates with the data of work[75]. The wide spectra of properties of cancerous cells are similar to the entangled (turbulent) structures with a wide range of values of DWH in the same crystal or amorphous alloy [28,30]. It is significant that the deformation activation of any genes produced by virus invasion, physical-chemical effects, physiological stresses with mutation formation, chronic inflammation and trauma, etc. demands the so-called threshold strain for malignant cell transformation (like it is for plastic flow in solids[2,17]) of not less than 5 mutations (the optimistic value is about 8 to 10 mutations in different cases). Therefore the next molecular-genetic events of replication cycles (the so-called self-activation of BC growth and evolution) generate the unique origin and composition of every tumor[77]. This means that the cancerous transformation of cells is not always controlled by the so-called oncogenes, but there is a general cause to regulate this process. As we can see it is the DWH of BC that is the same common reason for different effects like it was for the DWH in contrast to ROS only in aging and diseases[16]: the heavy DWH transforms the cells into the state with various diseases, aging and cancer[2,16]. Therefore the deformation drag of biological surfaces, cell membranes and organelles play the key role in their origin, evolution and transformation too. Similar explanations help to understand the periodical catastrophic multiplications of locust and plaque bacteria, etc. Malignant cells keep the possibility for apoptosis (fracture of cells[2]) or to retardation of proliferation in the same way as it is in the case of anticancer therapy under serious cell damage with high doses of irradiation, heavy drugs, high temperatures, etc.

There is some evidence for the phase transitions in stem cells which contribute to the origin of frequent prostate cancer in elderly men. The data of work on fruit flies (*Drosophila melanogaster*)[78] confirm this: the aging of embryonic stem cells (ESC) made their asymmetric fission, and tissue homeostasis has been weakened due to wrong orientation of stem cells, thus delaying the cell cycle. This cell aging is due to the DWH of their inter- and intracellular liquid and solid contents: in aged BC the deformation localizes mainly in primary directions like in solids[4,17] and liquids[70] under DWH, thus preventing ESC from right (useful) orientation at the cell cycle. Moreover, a paper from Mickie Bhatia's group was the first to directly confirm the heterogeneity of human embryonic stem cells (ESC)- their ability to develop into the pre-malignant ones and then to cancerous cells with small changes in chromosome structure[79]. This sub-population of human ESC lines was about 20 times more tumorigenic than the cultures they had been derived from and showed small changes in chromosome structure[79]. The close relationship between various subpopulations of pluripotent ESC proves the same origin and the key role of DWH in their transformations because after any phase transition the new phase are always heterogeneous like the anisotropy of DWH modes of plastic flow from the same source in solids[10,17]. Strong DWH makes the cells immortal like the long-living currents in superconductivity or continuous slip-climb of superfluid films of liquid $^4$HeII on the hard walls. Investigations confirm the key role of biological interfaces in the origin and proliferation of cancerous cells[80] as it is at phase transitions. This is in line with partial or complete spontaneous involution of malignant tumors in soft and slightly hardened tissues, young BC under soft physical-chemical stressing[2] like the rare reversible deformation in solids with low DWH [81], etc. The reversible deformation is possible under recoil of electrons in gases[82a], liquids and solids (negative resistance[82b]), gas molecules under light pressure[83], the abrupt and deep unloading[84], etc., and this effect is used for the softening therapy of BC in activation therapy[2].

The metabolism (mainly the electron fatigue deformation) of BC in the electron-transport-chains of mitochondria is determined by the direct DWH and the *reverse* DWS flow of electric currents. The DWS determines deformation durability of crystals and the longevity of BC due to the mechanical Bauschinger effect[2,9,33]. The data of work[85] strictly confirms this in BC by the facts that the less life span of 12 species (birds and mammals) is, the more the rate of $H_2O_2$-active oxidizing defects nucleation at the *reverse* electron transfer in mitochondria of cardiomyocytes. This means that it is the *lower reverse DWS* that makes lower concentration of $H_2O_2$ defects and the higher longevity of BC, and this is supported by the absence of higher life span at the *direct* electron transfer (the direct DWH of the Bauschinger effect)[2,39]. This is in line with the mechanism of superconductivity for various charges, superdiffusion in crystals under DWH up to nanostructures, superfluidity in various liquids, enhanced dissolving, corrosion or catalysis of deformation hardened solids, etc., which is always concerned with abrupt rise in DWH of materials[1].

The heavy cancer therapy is based on severe methods of tumor destruction (superheating, irradiation, physical-chemical effects, cell apoptosis stimulations, surgery, etc.). But the active physical treatments like the well-known numerous methods to destroy superconductivity or superfluidity (local acoustical deformation or non-thermal electromagnetic treatment of tumors with proper frequencies, etc.) or the faint methods concerned with DWS (small fractionated irradiation or heating/cooling and physical-chemical-physiological stressing/activation therapy, spiritual practices[2], increase of adaptation, DNA transformations, fitotherapy, homeopathy, etc.) and the combination of them are most prospective for anti-cancer therapy.

## 7. HYPERTONY, STROKE AND INFARCT

Intensive deformation prompts the phase transformations; the same is valid for the increased men's blood circulation under high blood pressure: vascular inflammation makes the DWH of BT at ultralow frequency of geomagnetic storms, high blood pressure and the excess of cholesterol in blood[15] assist the phase precipitation-plaque and thrombus nucleation in vessels, brain stroke and heart infarct.

## 8. EVOLUTION ROOTS OF LANGUAGES, CULTURE AND HISTORY[86].

Evolution origins of languages, culture and history are based on the deformation approach to the biological transformations during the long chain of events from DNA up to cells and BC growth. It was shown in[2], part 6 of this work that internal and external physiological stresses strengthen/soften the BC and change the activation of their phase transition (development and adaptation) from genes up to the growth of cells. All mental achievements (the origin of various languages, arts and music, culture and history, social behavior, etc.) are the direct products of this biological evolution, closely related social interlinking. Recent works reveal independent development of the same birds singing and the common variations among the calls of adult primates from the same region but containing consistent structural differences in calls between regions[87]. It may be suggested that typical agents of biological evolution – origin and aging, genetic activation, face/throat muscles and brain development, social factors and training, environmental effects, habitual and nasopharynx acoustics with different sounds being needed to best transmit information through different forest habitats, etc. can explain these coincidents and differences. Recent study of African genetic diversity has shown that Africans are descended from 14 ancestral populations, which often correlate with language and cultural groups[88]. The same transformations are happened to mental evolution, history[2] and human life, and it's corroborated by the same features and stages of the spontaneous and independent development, for example, in rock paintings of the late Stone Age and naïve art, the similar stages of some next arts and music developments in the different parts of the world, the same close stages of peoples' history and life. Recent mathematical investigations show that languages in Africa and Europe evolve in bursts like the species in evolutionary biology[89]. It is noteworthy that in some cases the development of the evolution tree of languages is very close to the parameters of biological evolution of the same human groups. Some of the indisputable confirmations to the above ideas are the facts that the biological neighborhood of European and African breeds promoted the invention of world-acknowledged jazz music in the United States, strong influence of African art on European art trends, and so on.

## 9. CONCLUSIONS

It is shown that laminar flow in liquids/plasma/gases is the equivalent to the movement of dislocation slip lines, bands and plastic shear in solids; the next stage of liquids deformation is turbulent flow– the multiplication and ordering of dislocation-like defects into subgrains and grain boundaries, then their rotation in the direction approximately perpendicular to the direction of shear flow, decrease in sizes of new structure modes[9], etc. Numerous proofs confirm the governing role of the same standard micromechanisms of plastic flow along the whole stress-strain curve and deformation at various phase transitions [2,39]. This approach helps to understand all physical-chemical effects at phase transitions and shows that the superconductivity, both quantum Hall effects are the typical cases of conventional quasi-continuous and jump-like deformation under current flow (for example, colossal expansion of zinc oxide nanowires after electronic excitation[90]). Liquid/plasma/gas flow effects (including superfluidity of solid/liquid He and ordinary liquids, ultradiffusion in deformation nanostructures, etc) are determined by the standard mechanisms of deformation too. Phase transformations in biological cells (explosive events of diversity and population of species and diseases – for example, locust and plaque bacteria, evolution, aging and cancer[2], bursts in the development of human intellectual possibilities (languages, culture, arts and sciences, history, etc.) depend on the same deformation effects in biological evolution.

---

.